\documentclass[a4paper,oneside,final,notitlepage,onecolumn]{article}
\usepackage{amsfonts}

\begin{document}

%%%%%%%%%%%%%%%%%%%%%%%%%%%%%%%%%%%%%%%%%%%%%%%%%%%%%%%%%%%%%%%%%%%%%%%%%%%%%%
\newtheorem{theorem}{Theorem}[section]
\newtheorem{lemma}{Lemma}[section]
\newtheorem{proposition}{Proposition}[section]
\newtheorem{corollary}{Corollary}[section]
\newtheorem{conjecture}{Conjecture}[section]
\newtheorem{example}{Example}[section]
\newtheorem{definition}{Definition}[section]
\newtheorem{remark}{Remark}[section]
\newtheorem{exercise}{Exercise}[section]
\newtheorem{axiom}{Axiom}[section]
%%%%%%%%%%%%%%%%%%%%%%%%%%%%%%%%%%%%%%%%%%%%%%%%%%%%%%%%%%%%%%%%%%%%%%%%%%%%%%
\renewcommand{\theequation}{\thesection.\arabic{equation}} 
% A fenti parancs atdefinialja az egyenleteket szamozo parancsot
%%%%%%%%%%%%%%%%%%%%%%%%%%%%%%%%%%%%%%%%%%%%%%%%%%%%%%%%%%%%%%%%%%%%%%%%%%%%%%

\author{Istv\'{a}n R\'{a}cz\thanks{%
~Fellow of the Japan Society for the Promotion of Science, on leave of absence
from MTA-KFKI Research Institute for Particle and Nuclear Physics, email: 
istvan@yukawa.kyoto-u.ac.jp} 
\\ %EndAName
Yukawa Institute for Theoretical Physics\\
Kyoto University, Kyoto 606-01, Japan}

\title{On the existence of Killing vector fields\thanks{%
Dedicated to the memory of Professor \'Agoston B\'aba}
}
\maketitle

\begin{abstract}

In covariant metric theories of coupled gravity-matter systems the necessary and 
sufficient conditions ensuring the existence of a Killing vector field are 
investigated. It is shown that the symmetries of initial data sets are preserved 
by the evolution of hyperbolic systems.          

\end{abstract}

\section{Introduction}
\setcounter{equation}{0}

In any physical theory those configurations which possess symmetry are distinguished. 
They represent, for instance, the equilibrium states of the underlying systems. 
It might also happen that the relevant field equations are far 
too complicated not allowing us to have the general but only certain symmetric 
solutions of them. Einstein's theory of gravity provides immediate examples for 
these possibilities. In addition, in black hole physics the asymptotic 
final state of the gravitational 
collapse of an isolated object is expected to be represented by a stationary black 
hole solution (see e.g. \cite{he,wald} for further details and references). 
It is also known that those solutions of Einstein's equations which 
possess symmetries do represent critical points of the phase space
in the linear stability problem, i.e. in the investigation of the 
validity of the perturbation theory \cite{vm1,vm2,ja1,ja2,ja&jem}. 

\smallskip

The above examples indicate that the identification of spacetimes possessing a symmetry 
is of obvious physical interest. In this paper the existence of symmetries will be 
studied in the case of coupled gravity-matter systems of the following types: The matter 
fields are assumed to satisfy suitable hyperbolic equations and to possess a minimal 
coupling to gravity. In addition, the Ricci tensor is supposed to be given as a function 
of the matter fields, their first covariant derivatives and the metric. 
Within this setting we address the following question: What are the necessary and 
sufficient conditions that can guarantee the existence of a Killing vector field 
so that the matter fields are also invariant?

\smallskip

The main purpose of the present paper is to answer this question. There is, 
however, a more pragmatic motivation beyond. We also would like to 
establish a detailed enough framework within which some of the claims 
of \cite{r} concerning the existence of Killing vector fields 
within the characteristic initial value problem in the case of 
Einstein--Klein-Gordon, Einstein--[non-Abelian] Higgs or 
Einstein-[Maxwell]-Yang-Mills-dilaton systems (see \cite{frw} for 
analogous results concerning the Einstein-Maxwell case) can be justified.

\smallskip 

Our main result is that hyperbolic evolutions preserve the symmetries of 
the initial data sets. In particular, initial symmetries are preserved in 
the case Einstein--Klein-Gordon, Einstein--[non-Abelian] Higgs and 
Einstein-[Maxwell]-Yang-Mills-dilaton systems.

\smallskip 

The necessary and sufficient conditions ensuring the existence of a 
Killing vector field are given in terms of restrictions on certain initial 
data sets in an initial value problem associated with the considered system. 
There is only one significant requirement the applied initial value problem 
has to satisfy: 
The existence and uniqueness of solutions to quasilinear wave equations 
is expected to be guaranteed within its framework in the smooth setting. 
The initial value problems which satisfy this condition will be referred as 
`appropriate initial value problems'. Immediate examples for appropriate 
initial value problems are the standard Cauchy problem 
(see e.g. Ref. \cite{ch}) and also in the 
characteristic initial value problem associated with an initial hypersurface 
represented by either the union of two smooth null hypersurfaces intersecting on a 
2-dimensional spacelike surface \cite{mzh,rendall} or a characteristic cone 
\cite{fried,cag,rendall}. Since no further requirement on the initial value problem 
is used anywhere in the derivation of our results we shall not make a definite choice 
among these appropriate initial value problems. However, as a direct consequence 
of this, our necessary and sufficient conditions 
cannot be as detailed as the corresponding restrictions applied e.g. in
\cite{bc} where the case of Einstein-Maxwell systems in the framework 
of the standard Cauchy problem was considered. Although, the analogous 
investigations could be carried out case by 
case they are not attempted to be done here. Instead we introduce the following 
notation that will be applied on equal footing to any of the above mentioned 
appropriate initial value formulations: The initial data sets will be represented by 
the basic variables in square brackets while the initial hyper surface will always be 
denoted by $\Sigma$. Accordingly, whenever an `initial data set' and an `initial 
hypersurface' will be referred the relevant pair of these notions defined in either of the 
`appropriate initial value problems' could be substituted.\footnote{%
All of the relevant field equations considered in this paper will (or in certain 
cases by making use of a `hyperbolic reduction procedure' \cite{fri1} will be 
shown to) possess the form 
of quasilinear wave equations. Thereby, in the particular case when our basic unknown 
is a function $\varphi$, one should think of the initial data set represented by 
$[\varphi]$ as a pair of the functions $\varphi\vert_\Sigma$ and 
$n^a\nabla_a\varphi\vert_\Sigma$, where $n^a$ is normal to $\Sigma$, in the standard 
Cauchy problem while it consists of a single function $\varphi\vert_\Sigma$ in the 
characteristic initial value problem.}

\smallskip

This paper is organized as follows: In section 2 we specify the gravity-matter systems
to which our main results apply. Section 3 starts with 
the construction of a `candidate' Killing vector field. Then a general 
procedure selecting the true Killing fields and also yielding the desired 
necessary and sufficient conditions is presented. Finally, in section 4 
the particular case of Einstein-[Maxwell]-Yang-Mills-dilaton systems is considered.

\smallskip

Throughout this paper a spacetime $(M,g_{ab})$ is taken to be a
smooth, paracompact, connected, orientable manifold $M$ endowed with a
smooth Lorentzian metric $g_{ab}$. Unless otherwise stated we shall use the 
notation and conventions of \cite{wald}.

\section{The gravity-matter systems}\label{g-m}
\setcounter{equation}{0}

This section is to give a mathematically precise specification of the 
considered coupled gravity-matter systems. 

The matter fields are considered to be represented by smooth tensor fields. Since the 
metric $g_{ab}$ is assumed to be defined everywhere on $M$ we may, without loss of 
generality, restrict considerations to tensor fields $T_{_{(i)}}{}_{a...b}$ of type 
$(0,l_i)$ where ${a...b}$ denotes the $l_i(\in \mathbb{N}\cup\{0\}$) `slots' of 
$T_{_{(i)}}{}_{a...b}$. The matter fields might also have gauge dependence
but, even if they have, the corresponding gauge or internal space indices 
will be suppressed. In most of the cases the spacetime indices of the
matter field variables will also be suppressed and the relevant $(0,l_i)$ type tensor 
fields will simply be denoted by $\mathcal{T}_{_{(i)}}$. All the indices, however, 
will be spelled out explicitly in each of the not self explaining situations.

In our setting, just like in various other analogous investigations, the hyperbolic 
character of the matter field equations have more relevance than any 
further information related, for instance, to their particular form. 
Thereby, we shall assume that the matter fields $\mathcal{T}_{_{(i)}}$
satisfy a quasi-linear, diagonal, second order hyperbolic system of the form
\footnote{%
To be able to derive our later results, see equations (\ref{evol1}) and 
(\ref{evol2}), we need to keep our formalism in an explicitly covariant form. 
As opposed to the usual treatment of the matter fields in the initial value 
problems (see e.g. \cite{wald,mzh,rendall}) the matter field equations are 
assumed to be tensor equations. This, in particular, means that instead of 
requiring that the components of matter field variables, as a set of 
functions, satisfy hyperbolic wave equations involving only the `flat wave 
operator' $g^{\mu\nu}\partial_\mu\partial_\nu$ exclusively (\ref{m1}) 
contains the complete action of the operator $g^{ab}\nabla_a\nabla_b$
on the matter field variables.}
\begin{equation}
\nabla^a\nabla_a\mathcal{T}_{_{(i)}}=\mathcal{F}_{_{(i)}}
\left(\mathcal{T}_{_{(j)}},\nabla_c\mathcal{T}_{_{(j)}},g_{ef}\right), 
\label{m1}
\end{equation}
where each of the $(0,l_i)$ type tensor fields $\mathcal{F}_{_{(i)}}$ is assumed to 
be a smooth function of the indicated arguments 
(the obvious dependence on the points of $M$ will be suppressed throughout this paper). 

We assume, furthermore, that the matter fields are coupled to gravity so that
the Ricci tensor $R_{ab}$ can be given as a smooth function of the fields 
$\mathcal{T}_{_{(i)}}$, their first covariant derivatives and the metric,
\begin{equation}
R_{ab}=R_{ab}\left(\mathcal{T}_{_{(i)}},\nabla_c\mathcal{T}_{_{(i)}},
g_{ef}\right).\label{g1}
\end{equation}
The last condition is satisfied, for instance, in Einstein's theory of gravity 
whenever the Lagrangian does contain at most first order derivatives of the matter 
field variables. Thereby, our results immediately apply to Einstein-matter systems of 
this kind. Note, however, that the above conditions are also satisfied
by the `conformally equivalent representation' of higher curvature theories possessing 
a gravitational Lagrangian that is a polynomial of the Ricci scalar (considered e.g. 
in Ref. \cite{jk}) and could also be satisfied by various other types of covariant 
metric theories of gravity.

\smallskip

The introduction of the above level of generality might seem to be pointless because 
even within Einstein's theory of gravity it has not been fixed in general whether 
such a coupled gravity-matter system possesses unique smooth solutions in suitable
initial value formulations for smooth initial data sets.
The remaining part of this section is to 
demonstrate -- by making use of a straightforward adaptation of the `hyperbolic 
reduction' procedure of Friedrich \cite{fri1} to the present case -- that (\ref{m1}) 
and (\ref{g1}) can be recast into the form of a system of coupled quasilinear wave 
equations and thereby they possess, up to diffeomorphisms, unique solutions 
in `appropriate initial value problems' for the smooth setting. 

Consider first the 
components of the metric $g_{ab}$ and the matter fields $\mathcal{T}_{_{(i)}}$ in some
coordinates as the basic unknowns. Observe then that the Ricci tensor and 
$\nabla^a\nabla_a\mathcal{T}_{_{(i)}}$ read in the associated local coordinates as
\begin{equation}
R_{\alpha\beta}=-\frac{1}{2}g^{\mu\nu}\partial_\mu\partial_\nu g_{\alpha\beta}+
g_{\delta(\alpha}\partial_{\beta)}\Gamma^\delta+H_{\alpha\beta}'(g_{\varepsilon\rho},
\partial_{\gamma}g_{\varepsilon\rho})
\end{equation} and
\begin{equation}
\nabla^\mu\nabla_\mu\mathcal{T}_{_{(i)}}=g^{\mu\nu}\partial_\mu\partial_\nu
\mathcal{T}_{_{(i)}}-\sum_{k=1}^{l_i}\left(\mathcal{T}_{_{(i)}}\right)
{}^{ [\alpha_k] }_{\delta}
\left({R_{\alpha_k}}^\delta+\partial_{\alpha_k}\Gamma^\delta\right)+
\mathcal{H}_{_{(i)}}'(g_{\varepsilon\rho},\partial_{\gamma}g_{\varepsilon\rho},
\mathcal{T}_{_{(j)}},\partial_{\gamma}\mathcal{T}_{_{(j)}}),
\end{equation}
where $\partial_{\alpha}$ denotes the partial derivative operator with respect to the
coordinate $x^\alpha$, $\Gamma^\mu=g^{\alpha\beta}{\Gamma^\mu}_{\alpha\beta}$ with
${\Gamma^\mu}_{\varepsilon\rho}$ denoting the Christoffel symbol of $g_{\alpha\beta}$,
moreover, $\left(\mathcal{T}_{_{(i)}}\right){}^{[\alpha_k]}_{\delta}$ stands for 
$T_{_{(i)}}{}_{\alpha_1...\alpha_{k-1}\delta\alpha_{k+1}...\alpha_{l_i}}$ and
$H_{\alpha\beta}'$ and $\mathcal{H}_{_{(i)}}'$ (the later with the suppressed
indices $\alpha_1...\alpha_{l_i}$) are appropriate smooth functions of the indicated
variables. Thereby, (\ref{m1}) and (\ref{g1}) can be recast into the form
\begin{equation}
g^{\mu\nu}\partial_\mu\partial_\nu\mathcal{T}_{_{(i)}}=\sum_{k=1}^{l_i}
\left(\mathcal{T}_{_{(i)}}\right){}^{[\alpha_k]}_{\delta}
\partial_{\alpha_k}\Gamma^\delta+
\mathcal{H}_{_{(i)}}(g_{\varepsilon\rho},\partial_{\gamma}g_{\varepsilon\rho},
\mathcal{T}_{_{(j)}},\partial_{\gamma}\mathcal{T}_{_{(j)}})\label{m2}
\end{equation}
\begin{equation}
g^{\mu\nu}\partial_\mu\partial_\nu g_{\alpha\beta}=
2g_{\delta(\alpha}\partial_{\beta)}\Gamma^\delta+H_{\alpha\beta}(g_{\varepsilon\rho},
\partial_{\gamma}g_{\varepsilon\rho},\mathcal{T}_{_{(j)}},
\partial_{\gamma}\mathcal{T}_{_{(j)}}).\label{g2}
\end{equation}
If the functions $\Gamma^\delta$ were known these equations would immediately give 
rise to a quasilinear, diagonal second order hyperbolic system for the unknowns and 
we would have done. Notice, however, that as in \cite{fri1} $\Gamma^\delta$ 
can be replaced in the above equations with arbitrary 
`gauge source functions' $f^\delta$ satisfying that $f^\delta=\Gamma^\delta$ on an 
arbitrarily chosen initial hypersurface $\Sigma$ (where $\Gamma^\delta$ is always 
determined by the specified initial data $[g_{\alpha\beta}]$) and 
consider the relevant set of `reduced equations' as propagation equations for the basic 
unknowns. From the corresponding unique solution the functions 
$\Gamma^\delta$ can be determined and, by refering to the twice contracted Bianchi 
identity, it can be demonstrated \cite{fri1} (see also page 540 of \cite{fri2}) that 
\begin{equation}
\nabla^\mu\nabla_\mu(\Gamma^\delta-f^\delta)+{R^\delta}_\nu(\Gamma^\nu-f^\nu)=0.
\label{s1}
\end{equation}
Recall then that $\Gamma^\delta=f^\delta$ on $\Sigma$ and by (\ref{g2}) and 
its reduced form the entire 
initial data set for (\ref{s1}) has to vanish there. Thus we have that $\Gamma^\delta=
f^\delta$ throughout the associated development, i.e. the solution of the reduced 
problem does satisfy the original set of coupled equations (\ref{m2}) and (\ref{g2}). 

\section{The construction of a Killing vector field}\label{sec-ckf}
\setcounter{equation}{0}

A spacetime admits a Killing vector field $K^a$ whenever the Killing equation
\begin{equation}
\mathcal{L}_K g_{ab}=\nabla_a K_b+\nabla_b K_a=0
\end{equation}
holds. Then we have that $X_{ab}=\nabla_a K_b$ is a 2-form field on $M$ and its 
integrability condition $(dX)_{abc}=0$ reads as
\begin{equation}
\nabla_a\nabla_b K_c+\nabla_c\nabla_a K_b+\nabla_b\nabla_c K_a=0.
\end{equation}
Replacing now $\nabla_a K_b$ with $-\nabla_b K_a$ (based on the antisymmetry of 
$\nabla_a K_b$) in the second term and using the definition of the Riemann tensor we
get that 
\begin{equation}
\nabla _a\nabla _bK_c+{R_{bca}}^dK_d=0,  \label{RK}
\end{equation}
is equivalent to the integrability condition of $X_{ab}$. This equation holds for 
any Killing vector field on $(M,g_{ab})$. We will show below that its contraction 
\begin{equation}
\nabla ^a\nabla _aK_c+{R_c}^dK_d=0,  \label{LK}
\end{equation}
which is a linear homogeneous wave equation for $K^a$, provides a mean to construct a 
`candidate' Killing vector field.  Clearly, any Killing vector field has to satisfy (\ref{LK}) 
but not all of its solutions will give rise to a Killing vector field. Our aim in this 
section to give the necessary and sufficient conditions on the initial data for 
(\ref{LK}), in terms of $g_{ab}$ and $\mathcal{T}_{_{(i)}}$, which guarantee that the 
corresponding solution of (\ref{LK}) will be a Killing vector field so that the matter 
fields will also be invariant, i.e. $\mathcal{L}_{K}\mathcal{T}_{_{(i)}}=0$.

The remaining part of this section is to show:

\begin{theorem}
Let $(M,g_{ab})$ be a spacetime associated with a gravity-matter system specified 
in section 2. Denote by $D[\Sigma]$ the domain of dependence of an initial hypersurface 
$\Sigma$ within an appropriate initial value problem. Then there exists a non-trivial 
Killing vector field $K^a$ on $D[\Sigma]$, with $\mathcal{L}_K\mathcal{T}_{_{(i)}}=0$, 
if and only if there exists a non-trivial initial data set $[K^a]$ for (\ref{LK}) so that 
$[\mathcal{L}_{K}{g}_{ab}]$ and $[\mathcal{L}_{K}\mathcal{T}_{_{(i)}}]$ vanish\footnote{%
As it will be shown below the `evolution equations' for $\mathcal{L}_{K}{g}_{ab}$ and 
$\mathcal{L}_{K}\mathcal{T}_{_{(i)}}$ are (\ref{evol1}) and (\ref{evol2}). By refering to
these equations one can make immediate sense of the last part of our condition in either 
of the above mentioned appropriate initial value problems.} identically on $\Sigma$.
\end{theorem}

\noindent
{\bf Proof}{\ } The necessity of the above conditions is trivial since the fields 
$\mathcal{L}_{K}{g}_{ab}$ and $\mathcal{L}_{K}\mathcal{T}_{_{(i)}}$, along with their
derivatives, vanish on $\Sigma$. 

\smallskip

To see that the above conditions are 
also sufficient one can proceed as follows: Suppose that $K^a$ satisfies (\ref{LK}) 
but otherwise it is an arbitrary vector field. It can be shown (by taking the
covariant derivative of (\ref{LK}), commuting derivatives and applying the contracted 
Bianchi identity) that $\mathcal{L}_{K}{g}_{ab}$ satisfies the equation
\begin{equation}
\nabla ^e\nabla _e\left(\mathcal{L}_Kg_{ab}\right)=
-2\mathcal{L}_KR_{ab}+2{{R^e}_{ab}}^{f}(\mathcal{L}_Kg_{ef})
+2R_{\;(a}^e(\mathcal{L}_Kg_{b)e}). \label{wsab}
\end{equation}
Moreover, by taking the Lie derivative of (\ref{g1}) we get\footnote{%
Whenever $T_{a_1...a_k}$ and $S_{b_1...b_l}$ are tensor 
fields of type $(0,k)$ and $(0,l)$, respectively,
$(\partial T_{a_1...a_k}/\partial S_{b_1...b_l})$ is considered 
to be a tensor field of type $(l,k)$. Accordingly, the contraction 
$(\partial T_{a_1...a_k}/\partial S_{b_1...b_l})\mathcal{L}_KS_{b_1...b_l}$ 
of $(\partial T_{a_1...a_k}/\partial S_{b_1...b_l})$ and 
$\mathcal{L}_KS_{b_1...b_l}$ is a tensor field of type $(0,k)$.}
\begin{equation}
\mathcal{L}_KR_{ab}=\sum_{(i)}\left(\frac{\partial R_{ab}}{\partial 
\mathcal{T}_{_{(i)}}}\right)\mathcal{L}_K\mathcal{T}_{_{(i)}}
+\sum_{(i)}\left(
\frac{\partial R_{ab}}{\partial\left(\nabla_e\mathcal{T}_{_{(i)}}\right)
}\right)\mathcal{L}_K\left(\nabla_e\mathcal{T}_{_{(i)}}\right)+\left(
\frac{\partial R_{ab}}{\partial g_{ef}}\right)\mathcal{L}_Kg_{ef}. \label{lie1}
\end{equation}
Consider now the commutation relation of $\mathcal{L}_K$ and $\nabla_a$ 
\begin{equation}
\mathcal{L}_K\left(\nabla_b{\mathcal{T}_{_{(i)}}}\right)=
\nabla_b\left(\mathcal{L}_K{\mathcal{T}_{_{(i)}}}\right)
-\sum_{k=1}^{l_i} \left(\mathcal{T}_{_{(i)}}\right){}^{[a_k]}_e
\left[\nabla\mathcal{L}_Kg\right]{{{}_{a_k}}^e}_{b},\label{com}
\end{equation}
where $\left(\mathcal{T}_{_{(i)}}\right){}^{[a_k]}_e$ stands for 
$T_{_{(i)}}{}_{a_1...a_{k-1}ea_{k+1}...a_{l_i}}$ and the notation 
\begin{equation}
\left[\nabla\mathcal{L}_Kg\right]{{{}_{a}}^c}_{b}=\frac{1}{2}g^{cf}
\left\{\nabla_{a}\left(\mathcal{L}_Kg_{fb}\right)+
\nabla_b\left(\mathcal{L}_Kg_{{a}f}\right)
-\nabla_f\left(\mathcal{L}_Kg_{{a}b}\right)\right\}
\end{equation}
has been used. Then, in virtue of (\ref{wsab}),
(\ref{lie1}) and (\ref{com}), $\mathcal{L}_Kg_{ab}$ satisfies an equation of the form
\begin{equation}
\nabla ^e\nabla _e\left(\mathcal{L}_Kg_{ab}\right)=
\mathrm{K}_{ab}(\mathcal{L}_Kg_{cd})+\mathrm{L}_{ab}\left(\nabla_c(\mathcal{L}_Kg_{cd})\right)
+\sum_{(i)}{\mathrm{M}_{_{(i)}}}{}_{ab}\left(\mathcal{L}_K\mathcal{T}_{_{(i)}}\right)
+\sum_{(i)}{\mathrm{N}{}_{_{(i)}}}{}_{ab}\left(\nabla_c(\mathcal{L}_K\mathcal{T}_{_{(i)}})
\right)\label{evol1}
\end{equation}
where $\mathrm{K}_{ab}$, $\mathrm{L}_{ab}$, ${\mathrm{M}_{_{(i)}}}{}_{ab}$ and 
${\mathrm{N}{}_{_{(i)}}}{}_{ab}$ are linear and homogeneous functions in their indicated 
arguments.

\smallskip

Now we show that the same type of equation holds for $\mathcal{L}_K\mathcal{T}_{_{(i)}}$.
To see this consider first the Lie derivative of (\ref{m1}) with respect to the 
vector field $K^a$ 
\begin{eqnarray}
\mathcal{L}_K\left(\nabla^e\nabla_e \mathcal{T}_{_{(i)}}\right)&=&
\left(\mathcal{L}_Kg^{ef}\right)\nabla_e\nabla_f \mathcal{T}_{_{(i)}}+
g^{ef}\mathcal{L}_K\left(\nabla_e\nabla_f\mathcal{T}_{_{(i)}}\right)
\label{lie2}\\ &=&
\sum_{(j)}\left(\frac{\partial \mathcal{F}_{_{(i)}}}
{\partial\mathcal{T}_{_{(j)}}}\right)
\mathcal{L}_K\mathcal{T}_{_{(j)}}
+\sum_{(j)}\left(\frac{\partial\mathcal{F}_{_{(i)}}}
{\partial\left(\nabla_c\mathcal{T}_{_{(j)}}\right)}\right)
\mathcal{L}_K\left(\nabla_c\mathcal{T}_{_{(j)}}\right)+
\left(\frac{\partial \mathcal{F}_{_{(i)}}}{\partial g_{ab}}\right)\mathcal{L}_Kg_{ab}.  
\nonumber
\end{eqnarray}
By applying the commutation relation (\ref{com}) twice we get 
\begin{eqnarray}
\mathcal{L}_K\left(\nabla_e\nabla_f\mathcal{T}_{_{(i)}}\right)&=&
\nabla_e\nabla_f\left(\mathcal{L}_K\mathcal{T}_{_{(i)}}\right)-
\left(\nabla_c\mathcal{T}_{_{(i)}}\right)
\left[\nabla\mathcal{L}_Kg\right]{{}_{f}}^c{}_e \nonumber
\\ &-&\sum_{k=1}^{l_i}\left[2\left(\nabla_{(e|}\mathcal{T}_{_{(i)}}\right){}^{[a_k]}_c
\left[\nabla\mathcal{L}_Kg\right]{{{}_{a_k}}^c}_{|f)}+
\left(\mathcal{T}_{_{(i)}}\right){}^{[a_k]}_c
\nabla_{e}\left(\left[\nabla\mathcal{L}_Kg\right]{{{}_{a_k}}^c}_{f}\right)\right].
\label{nt}
\end{eqnarray}

The last term on the r.h.s. of (\ref{nt}) does not possess the needed form yet because
it contains second order covariant derivatives of $\mathcal{L}_Kg_{{a}b}$. Thereby it 
could `get in the way' of the demonstration that $\mathcal{L}_K\mathcal{T}_{_{(i)}}$ 
do satisfy quasilinear wave equations of the type (\ref{evol1}). Note, however, that by the 
substitution $\mathcal{L}_Kg_{{a}b}=\nabla_aK_b+\nabla_bK_a$, along with the commutation 
covariant derivatives and the application of the Bianchi identities, we get that whenever
$K^a$ satisfies (\ref{LK})
\begin{equation}
g^{ef}\nabla_{e}\left(\left[\nabla\mathcal{L}_Kg\right]{{{}_{a_k}}^c}_{f}\right)=
\frac{1}{2}g^{cd}\left\{\nabla^e\nabla_e\left(\mathcal{L}_Kg_{a_kd}\right)-{R_d}^{e}
\left(\mathcal{L}_Kg_{a_ke}\right)+{R_{a_k}}^e\left(\mathcal{L}_Kg_{ed}\right)\right\}
\label{nt2}
\end{equation}
must also holds. Finally, by making use of all of the equations (\ref{lie1})--(\ref{nt2}), 
it can be justified that $\mathcal{L}_K\mathcal{T}_{_{(i)}}$ do really satisfy an equation
of the form
\begin{equation}
\nabla ^e\nabla _e\left(\mathcal{L}_K\mathcal{T}_{_{(i)}}\right)=
\mathcal{P}_{_{(i)}}\left(\mathcal{L}_Kg_{cd}\right)+
\mathcal{Q}_{_{(i)}}\left(\nabla_b(\mathcal{L}_Kg_{cd})\right)
+\sum_{(j)} \mathcal{R}_{_{(i)(j)}}\left(\mathcal{L}_K\mathcal{T}_{_{(j)}}\right)
+\sum_{(j)} \mathcal{S}_{_{(i)(j)}}\left(\nabla_c(\mathcal{L}_K\mathcal{T}_{_{(j)}})\right)
\label{evol2}
\end{equation}
where $\mathcal{P}_{_{(i)}}$, $\mathcal{Q}_{_{(i)}}$, $\mathcal{R}_{_{(i)(j)}}$ and 
$\mathcal{S}_{_{(i)(j)}}$ are linear and homogeneous functions in their indicated arguments.

\smallskip

To complete our proof consider a non-trivial initial data set $[K^a]$ on an initial hypersurface 
$\Sigma$ associated with vanishing initial data, $[\mathcal{L}_Kg_{ab}]$ and 
$[\mathcal{L}_K\mathcal{T}_{_{(i)}}]$ for (\ref{evol1}) and (\ref{evol2}). Since (\ref{evol1}) 
and (\ref{evol2}) comprise a set of coupled linear and homogeneous wave equations 
for $\mathcal{L}_Kg_{ab}$ and $\mathcal{L}_K\mathcal{T}_{_{(i)}}$, which
possesses the identically zero solution associated with a zero initial data set, we 
have that $\mathcal{L}_Kg_{ab}\equiv 0$ and $\mathcal{L}_K\mathcal{T}_{_{(i)}}\equiv 0$
throughout the domain where the associated unique solution of (\ref{LK}) does exist. 
To see, finally, that this domain has to coincide with $D[\Sigma]$ note that since (\ref{LK}) 
is also a linear homogeneous wave equation any solution of it can be shown -- by making
use of the `patching together local solutions' techniques described e.g. on page 266
of \cite{mzhs} -- to extend to the entire of the domain of dependence $D[\Sigma]$ of 
$\Sigma$. \hfill\fbox{}

\bigskip

Note that the full set of equations (\ref{m1}), (\ref{g1}), (\ref{LK}), (\ref{evol1}) 
and (\ref{evol2}) could also be recast -- by making use of the hyperbolic reduction 
techniques -- into the form of a set of quasilinear wave equations. Moreover, the 
validity of the conditions of the above theorem can also be justified if the 
initial data $[{g}_{ab}]$ and $[\mathcal{T}_{_{(i)}}]$, associated with (\ref{m1}) 
and (\ref{g1}), are given on $\Sigma$. Therefore in virtue of the above result we have 

\begin{corollary}
Denote by $D[\Sigma]$ the maximal development of an initial data set $[{g}_{ab}]$ and 
$[\mathcal{T}_{_{(i)}}]$, associated with (\ref{m1}) and (\ref{g1}), specified on an 
initial hypersurface $\Sigma$ within an appropriate initial value problem. 
Then there exists a non-trivial 
Killing vector field $K^a$ on $D[\Sigma]$, with $\mathcal{L}_K\mathcal{T}_{_{(i)}}=0$, 
if and only if there exists a non-trivial initial data set $[K^a]$ for (\ref{LK}) so that 
the initial data, $[\mathcal{L}_{K}{g}_{ab}]$ and $[\mathcal{L}_{K}\mathcal{T}_{_{(i)}}]$, 
for (\ref{evol1}) and (\ref{evol2}) vanish identically on $\Sigma$.
\end{corollary}

\begin{remark}
Remember that whenever only the existence of a Killing vector field ensured for the 
considered gravity-matter systems the matter fields need not to be invariant. 
Nevertheless, the non-invariance of them -- represented by the fields 
$\mathcal{L}_K \mathcal{T}_{_{(i)}}$ -- has to `develop' according to the linear 
homogeneous wave equations 
\begin{equation}
\nabla^e\nabla_e\left(\mathcal{L}_K \mathcal{T}_{_{(i)}}\right)=
\sum_{(j)}\left(\frac{\partial \mathcal{F}_{_{(i)}}}
{\partial\mathcal{T}_{_{(j)}}}\right)\mathcal{L}_K\mathcal{T}_{_{(j)}}
+\sum_{(j)}\left(\frac{\partial\mathcal{F}_{_{(i)}}}
{\partial\left(\nabla_e\mathcal{T}_{_{(j)}}\right)}\right)
\nabla_e\left(\mathcal{L}_K\mathcal{T}_{_{(j)}}\right) \label{lie1r} 
\end{equation}
and the constraints 
\begin{equation}
0=\sum_{(i)}\left(\frac{\partial R_{ab}}{\partial 
\mathcal{T}_{_{(i)}}}\right)\mathcal{L}_K\mathcal{T}_{_{(i)}}
+\sum_{(i)}\left(
\frac{\partial R_{ab}}{\partial\left(\nabla_e\mathcal{T}_{_{(i)}}\right)
}\right)\nabla_e\left(\mathcal{L}_K\mathcal{T}_{_{(i)}}\right), \label{lie2r}
\end{equation}
have also to be satisfied.
\end{remark}

\section{Einstein-[Maxwell]-Yang-Mills (E[M]YM) systems}\label{sec-EMYM}
\setcounter{equation}{0}

In this section the case of matter fields for which the r.h.s. of (\ref{m1}) explicitly 
depends on second order derivatives of the matter field variables will be considered. 
However, instead of trying to have a result of `maximal generality' we only 
illustrate what sort of modifications of the procedures of the previous section are needed 
to cover the particular case of E[M]YM systems
which are described as follows.

A Yang-Mills field is represented by a vector potential $A_a$ taking values in the Lie algebra $\mathfrak{g}$ 
of a Lie group $\mathrm{G}$. For the sake of definiteness, throughout this paper, 
$\mathrm{G}$ will be assumed to be a matrix group and it will also be assumed 
that there exists a real inner product, denoted by 
$(\ /\ )$, on $\mathfrak{g}$ which is invariant under the adjoint representation\footnote{%
Note that the inner product $(\ /\ )$ is not needed to be assumed to be positive definite 
as opposed to its use in \cite{r}.}. In terms 
of the gauge potential $A_a$ the Lie-algebra-valued 2-form field $F_{ab}$ is given as 
\begin{equation}
F_{ab}=\nabla_aA_b-\nabla_bA_a+
\left[A_a,A_b\right]\label{ymf}
\end{equation}
where $[\ ,\ ]$ denotes the product in $\mathfrak{g}$.
The Ricci tensor of an E[M]YM system is
\begin{equation}
R_{ab}=\chi\left\{\left(F_{ae}/{F_{b}}^{e}\right)
-\frac{1}{4}g_{ab}\left(F_{ef}/F^{ef}\right)\right\},\label{ymt}
\end{equation} 
with some constant $\chi$, while the Yang-Mills field equations read as
\begin{equation}
\nabla^aF_{ab}+\left[A^a,F_{ba}\right]=0.\label{yme}
\end{equation}

By the substitution of the r.h.s. of (\ref{ymf}) for $F_{ab}$ into (\ref{yme}) and 
commuting derivatives we get 
\begin{equation}
\nabla^a\nabla_a A_{b}={R_b}^dA_d-[\nabla^aA_a,A_b]-[A_a,\nabla^aA_b]-[A^a,F_{ba}]+
\nabla_b(\nabla^a A_a).\label{ymeq}
\end{equation}
In virtue of (\ref{ymf}), (\ref{yme}) and (\ref{ymeq}) we have 
that the considered E[M]YM system is almost of the type characterized in section 2. The 
only, however, significant difference is that on the r.h.s. of (\ref{ymeq}) we have the 
term $\nabla_b(\nabla^a A_a)$ which contains second order covariant derivatives of the 
gauge potential $A_a$. Fortunately, this term is of the same character as 
$\partial_\alpha\Gamma^\beta$ was in (\ref{g2}) and the same type of 
resolution does apply here. To see this substitute 
an arbitrary function $\mathcal{A}$, called `gauge source function', for $\nabla^a A_a$ in 
the last term of (\ref{ymeq}) so that $\mathcal{A}=\nabla^a A_a$ holds on an 
initial hypersurface
$\Sigma$. The yielded equation is referred to be the reduced form 
of (\ref{ymeq}). Then by the application of $\nabla^b$ to (\ref{ymeq}), 
and also to its reduced form, we get that 
\begin{equation}
\nabla^b\nabla_b (\mathcal{A}-\nabla^a A_a)=0.\label{s2}
\end{equation}
In addition, we have that $\mathcal{A}=\nabla^a A_a$ on $\Sigma$, moreover, (\ref{ymeq}), 
along with its reduced form, implies that the initial data for (\ref{s2}) has to vanish
identically on $\Sigma$. Thereby the unique solution of the reduced form of  (\ref{ymeq})
is in fact a solution of  (\ref{ymeq}) itself.

It follows from the above argument that an E[M]YM system differs from the systems specified 
in section 2 only in that in the present case we have to apply two types of gauge source 
functions $f^\delta$ and $\mathcal{A}$ and then use the subsidiary equations (\ref{s1}) 
and (\ref{s2}) to demonstrate that the unique smooth solutions of the `reduced equations' -- 
which equations are obviously the needed type of quasilinear wave equations -- 
are, up to `gauge transformations', unique solutions of the full E[M]YM equations, as well.

\medskip

Turning back to the problem of the existence of a Killing vector field in the case of 
an E[M]YM system consider first a vector field $K^a$ satisfying (\ref{LK}) and suppose
that $\mathcal{L}_K(\nabla^a A_a)$ does vanish throughout. Then, from (\ref{ymeq}) we 
get that an equation of the type (\ref{evol2}) has to be satisfied by 
$\mathcal{L}_K A_a$. Thereby, a statement analogous to that of Theor.3.1 can immediately 
be recovered in the case of E[M]YM systems whenever a gauge potential 
$A_a$ with $\mathcal{L}_K(\nabla^a A_a)=0$ does exist.

Suppose now that a non-trivial initial data set $[K^a]$ for (\ref{LK}) is given so that 
$[\mathcal{L}_Kg_{ab}]$ and $[\mathcal{L}_KA_a]$ vanish on an initial hypersurface 
$\Sigma$.  We show that then there exists a gauge potential $A^*_a$ (at least in a 
sufficiently small neighbourhood $\mathcal{O}$ of $\Sigma$) so that $[A^*_a]$ coincides 
with $[A_a]$ on $\Sigma$ and $\mathcal{L}_K(\nabla^a A^*_a)=0$. To see
this replace $A_a$ by a gauge equivalent potential
\begin{equation}
A^*_a=m^{-1}\left(\nabla_a m+A_am\right),\label{ymgau}
\end{equation}
where $m$ is an arbitrary smooth $\mathrm{G}$-valued function (defined in general locally 
on an open subset of $M$). Since 
$\mathcal{L}_K(\nabla^aA_a)=0$ on $\Sigma$ it is reasonable to require 
$[A^*_a]$ to coincide with $[A_a]$ on $\Sigma$. Accordingly, we set 
$m=\mathbb{E}$ (and if it is needed $\nabla_a m=0$) on $\Sigma$, where 
$\mathbb{E}$ is the unit element of $G$. 
Since, so far only the initial data $[A^*_a]$ has been specified, we still have the 
freedom to choose the gauge source function $\mathcal{A}^*=\nabla^aA^*_a$ arbitrarily. Thereby
we assume that $\nabla^aA^*_a$ is so that $\mathcal{L}_K(\nabla^aA^*_a)$ 
vanishes identically in $D[\Sigma]$. In addition, the generator of the 
gauge transformation (\ref{ymgau}) has to satisfy
\begin{equation}
\nabla^a\nabla_am=m(\nabla^aA^*_a)-(\nabla^aA_a)m-A_a\nabla^am+
(\nabla^am)m^{-1}(\nabla_am+A_am). \label{gtr}
\end{equation}
By making use of this equation $m$ and, in turn, the desired gauge potential $A^*_a$ can 
be constructed. Clearly, (\ref{gtr}), with given `source terms' $\nabla^aA^*_a$ 
and $\nabla^aA_a$, possesses (at least in a sufficiently small neighbourhood $\mathcal{O}$
of the initial hypersurface $\Sigma$) a unique solution so that $m=\mathbb{E}$ (and 
$\nabla_a m=0$) on $\Sigma$. Moreover, for the gauge potential $A^*_a$, determined by 
this unique solution $m$ via (\ref{ymgau}), $[\mathcal{L}_KA^*_a]$ vanishes on $\Sigma$ 
and $\mathcal{L}_K(\nabla^aA^*_a)=0$ throughout $\mathcal{O}$.

\medskip

The following sums up what have been proven

\begin{theorem}
Let $(M,g_{ab})$ be a spacetime associated with an E[M]YM system as it was specified 
above. Let, furthermore, $\Sigma$ be an initial hypersurface within an appropriate 
initial value problem. Then there exist a non-trivial Killing vector field $K^a$
and a gauge potential $A^*_a$ so that $\mathcal{L}_KA^*_a=0$ in a neighbourhood 
$\mathcal{O}$ of $\Sigma$, if and only if there exists a non-trivial 
initial data set $[K^a]$ for (\ref{LK}) so that $[\mathcal{L}_{K}{g}_{ab}]$ and 
$[\mathcal{L}_{K}A_a]$ vanish identically on $\Sigma$.
\end{theorem}

\begin{remark}
Combining now all the above results a statement analogous to that of Theor.4.1 can be 
proven 
for appropriate couplings (based on terms containing sufficiently low order derivatives
of the matter fields variables) of Yang-Mills -- matter systems in 
Einstein theory of gravity. In particular, it can be shown that Theor.4.1 generalizes to 
the case of Einstein-[Maxwell]-Yang-Mills-dilaton systems given by the Lagrangian
\begin{equation}
\mathcal{L}=R+2\nabla^e\psi\nabla_e\psi-e^{2\gamma_{_{d}}\psi}(F_{ef}/F^{ef}),
\end{equation}
where $R$ is the Ricci scalar, $\psi$ denotes the (real) dilaton field and 
$\gamma_{_{d}}$ is the dilaton coupling constant.
\end{remark}

\section*{Acknowledgments}
I wish to thank Helmut Friedrich for stimulating discussions.
This research was supported in parts by the Monbusho Grant-in-aid No. P96369 and by the
OTKA grant No. T030374. 

\vfill\eject

\end{document}